# Photovoltaic (PV) Virtual Inertia and Fast Frequency Regulation in High PV Power Grids

Shutang You

*Abstract*— This paper studies the frequency response using PV. Multiple control strategies are considered and simulated in the high PV ERCOT model, including inertia control, synthetic governor control, and AGC control. The impact of different parameters in PV inertia control and their correlation and impact on frequency response are analyzed. The simulation results show that PV farm has potential to provide multiple types of grid service to support system frequency. This paper also proposed a distributed fast frequency control approach that can better leverage the PV headroom reserve to improve the system frequency nadir after contingencies.

## I. INTRODUCTION

Due to price decrease, Photovoltaic (PV) and wind power generation penetration is increasing many power systems. Power grid reliability is highly important due to its vast impact areas. Due to the unique features of renewable generation, different approaches have been developed to better understand the system dynamics and study methods to mitigate their impacts. Similar to other renewable generation, PV usually runs at the maximum power point, providing no frequency response to the power grid. The displacement of synchronous generators with PV has direct impacts on the system inertia level and frequency regulation capability. Many power systems noticed the risks of insufficient system inertia and frequency regulation resources due to the increase of PV and other non-synchronous renewable generation. System frequency response focuses on whether the system can prevent frequency variation from triggering protection relays subject to resource contingencies, including power unit/plant trips and sudden load increases in normal operation and extreme weather conditions.

Ref. [1]studied the frequency response of South Australia power grid with high penetration of renewable generation, and showed that insufficient inertia will influence frequency regulation. In Ref. [2], the Irish power grid faces challenges in operating at the 50% penetration rate of wind because of reduced inertia. After reviewing the studies on several power grids with increasing PV and wind penetrations, Ref. [3] found that additional control strategies and resources are needed to meet the primary frequency control demand as renewable penetration increases. United States has substantial solar energy resource [4]. The SunShot Initiative goal predicts that solar energy has the potential to generate 14% of the total electricity consumption in the U.S. by 2030, and this percentage will increase to 27% by 2050 [4]. Some preliminary studies in the U.S. power grids demonstrated that the overall frequency response will decrease significantly with renewable penetration increase [5, 6].

## II. REVIEW ON PV FREQUENCY CONTROL

### A. PV Frequency Droop Control

PV frequency droop control (primarily for overfrequency regulation) has become a standard in North America power grids. The NERC reliability guideline on BPS-connected inverter-based resource performance [7] and the latest IEEE 1547 standard (expected to be published in 2019) require that smart inverters provide frequency-watt function to decrease real power to stabilize over-frequency events. If active power is available, inverters should also increase real power to support low frequency.

The NERC reliability guideline specifies that "the active power-frequency control system should have an adjustable proportional droop characteristic with a default value of 5%". In addition, frequency droop should be based on the difference between maximum nameplate active power output (Pmax) and zero output (Pmin) such that the droop line is always constant for a resource. The IEEE 1547 standard does not have explicitly requirement on the droop characteristics.

The NERC guideline also has requirements on the dynamic active power-frequency performance. It specifies values of the reaction time, rise time, settling time, overshoot, and settling band after a 0.002 p.u. (0.12Hz) step change in frequency from nominal 60Hz.

### B. PV Synthetic Inertia Control

As of 2018, North American power grids have no standard or requirement on synthetic inertia control for inverter-based resources. However, it is widely accepted that synthetic inertia has two desired functions: keeping the initial RoCoF value below the maximum capability of generators and loads to remain connected; and limit the frequency nadir to avoid load/generator disconnection. While few studies have studied PV synthetic inertia control, most inverter inertia control studies focus on wind generation, EV (energy storage), HVDC, and virtual synchronous generators (VSG).

Studies on inertia control using EV, HVDC, and VSG use the inverter real power control loop to mimic the inertia response of conventional synchronous generators. High-pass

This work is funded in whole by the U.S. Department of Energy Solar Energy Technologies Office, under Award Number 30844. This work also made use of Engineering Research Center Shared Facilities supported by the Engineering Research Center Program of the National Science Foundation and the U.S. Department of Energy under NSF Award Number EEC-1041877 and the CURENT Industry Partnership Program.

Shutang You is with the Department of Electrical Engineering and Computer Science, the University of Tennessee, Knoxville, TN 37996 USA (e-mail: syou3@utk.edu).

filtered frequency signal and df/dt are the two common inputs for inertia control. For example, Ref. [8] studied virtual inertia emulation in wind turbine generator inverters and used the high-pass filtered frequency signal to control electrical output. Ref. [3] applied df/dt in control to provide inertia response from HVDC-interfaced wind power.

Existing studies on synthetic inertia find inverter-based inertia emulation has some characteristics that are different from the rotor inertia of conventional synchronous generators. These characteristics are briefly categorized as follows.

*The need for synthetic inertia:* The need for synthetic inertia applies for small synchronous areas with high penetration of renewables and large synchronous areas to prevent total system collapse in case of a system split and subsequent islanding operation [9]. For example, in Hydro Quebec, Wind Generation Resources (>10 MW) during underfrequency conditions are required to provide momentary overproduction that limits the frequency drop after a major loss of generation.

*Stability*: Ref. [10]found that inertia controllers need to be designed individually for each power grid to avoid instability caused by the limited bandwidth of df/dt.

*Response speed (reaction time and rise time):* Ref. [8] found that the "synthetic inertia" fast frequency response (FFR) type devices have the potential to prevent high RoCoF events but the time period required to reliably detect and measure RoCoF events to ensure the appropriate response to mitigate the events poses some challenges. Ref. [11] found that the synthetic inertia needs to satisfy serval criteria to maintain RoCoF within a specific range ($\pm 0.5$ Hz/s for the Ireland power grid). 1) a short reaction time (begin responding within 100ms from the start of the event for the Ireland grid); 2) ramp at a sufficient rate to deliver power to the system. (Active power injection must be fully achieved 200 milliseconds after the device begins to respond for the Ireland grid);

*Response characteristics during frequency recovery:* The study in [11]found that a suitable form of synthetic inertia control is needed to prevent unintended adverse system issues during the frequency recovery. Recognizing that frequency recovery does not inertia support, Ref. [12]found that compared with fixed inertia of conventional generators, emulating varying inertia may help improve the system frequency stability. A study in Ref. [13] showed that some virtual inertia controls that need to be paid back (for example, wind generation) may not always be beneficial to system frequency response, especially when the system governor response is fast.

*Response power magnitude:* Ref. [11] found that a minimum response power is needed to constrain RoCoF ($\pm 360$ MW of supplementary synthetic inertia for the Ireland grid would need to be available for the duration of the RoCoF event).

*Respond to both high and low frequency events*: Ref. [11] found that Synthetic inertia response is required for both high and low frequency *events.*

This paper will study the characteristics of solar inertia control, frequency droop control, and AGC control, as well as their implementation in the U.S. ERCOT system. The study results will provide a guideline on implementing PV frequency control in high-PV low-inertia power grids.

### III. SYNTHETIC INERTIA CONTROL OF PV

PV synthetic inertia uses the PLL frequency as input. Its control diagram is shown in Figure 1. It uses a deadband, a low pass filter, a control gain, and a differential link, and a power limit link to accomplish the inertia emulation function. Different from inertia of conventional synchronous generators, the PV synthetic inertia has time delay and will be limited by the available power reserve of the inverter.

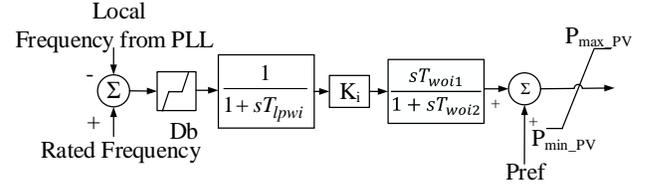

Figure 1. Synthetic inertia control using PV

To better determine PV virtual inertia control parameters, it is desirable to study the relation between control parameters and the virtual inertia contribution from PV inverters. As seen from simulation results, the primary function of PV virtual inertia control is injecting electric power when the system frequency declines. The direct input for PV virtual inertia control is the rate of change of frequency (ROCOF). As the unit step function is one of the most common test inputs to examine the response of a control system, it is straightforward to consider using a step function of ROCOF as the input to study the virtual inertia controller. Moreover, a step function of ROCOF is also very close to the system actual measurement the initial stage of a generation loss contingency, since the system frequency decline rate is a constant right after a system frequency contingency and before the governor response and load damping take effects. Thus, the rate of change of frequency can be obtained by (1).

$$ROCOF = \frac{P_{imbalance}/C_{system}}{2H_{System}} \cdot f_N \quad (1)$$

where $P_{imbalance}$ is the power imbalance magnitude (MW). $C_{system}$ is the total system capacity (MVA). $H_{System}$ is the system inertia time constant (second). $f_N$ is the system nominal frequency (Hz). In (1), using the inverter virtual inertia power ($P_{inv\_inertia}$) to replace power imbalance ($P_{imbalance}$), and the inverter capacity ($C_{system}$) to replace the system capacity ($C_{inv}$), the inverter inertia time constant can be similarly defined as the virtual inertia response power value with respect to the inverter capacity for a certain rate of change of frequency.

$$H_{inv} = \frac{P_{inv\_inertia}/C_{inv}}{2 \cdot ROCOF/f_N} \quad (2)$$

Usually, considering the response dynamics of in the virtual inertia controller, the inverter inertia output can be a time-variant value:

$$p_{inv\_inertia} = p_{inv\_inertia}(t) \quad (3)$$

Substitute (3) into (2), the instantaneous virtual inertia time constant can be defined as:

$$H_{inv}(t) = \frac{f_N}{2 \cdot ROCOF \cdot C_{inv}} \cdot p_{inv\_inertia}(t) \quad (4)$$

One example of inverter inertia control response is to use a unit step function in ROCOF. Figure 2 shows the frequency input profile with a step in ROCOF. The synthetic inertia power of PV is shown in Figure 3.

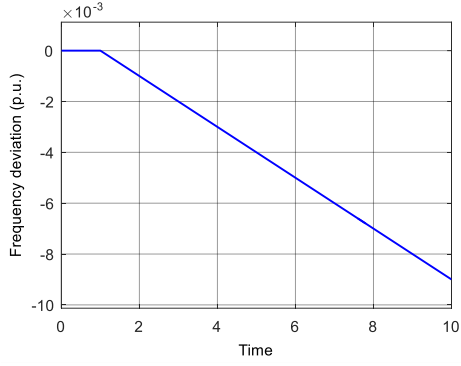

Figure 2. Frequency deviation in per unit value as input to the PV virtual inertia controller

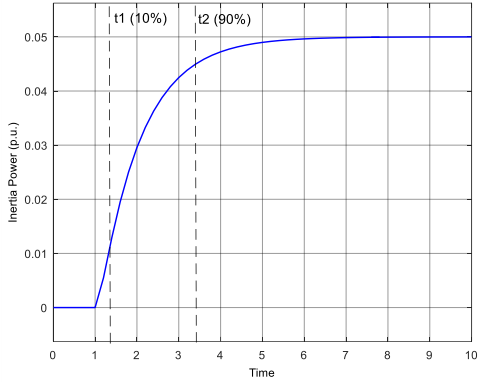

Figure 3. PV inverter virtual inertia response output

From Figure 3, it can be seen that inertia power output increases from 0 (the initial value) to 0.05 per unit (the steady-state value). Three metrics can be defined to quantify the characteristics of the virtual inertia response of a PV inverter.

### A. Steady-state virtual inertia time constant (second).

The steady-state virtual inertia time constant is equivalent to the inertia time constant of a synchronous machine. The steady-state virtual inertia time constant can be obtained using a step function of ROCOF and examine the steady-state response of the inverter output. In the control diagram shown in Figure 2, the steady-state virtual inertia time constant is also determined by the control gain and time constant of the low pass filter.

$$H_{inv\_steady\_state} = \frac{f_N}{2 \cdot ROCOF \cdot C_{inv}} \cdot p_{inv\_steady\_state} = \frac{K_i \cdot T_{woi1}}{2} \quad (5)$$

### B. 10%-90% open-loop virtual inertia response rise time (second)

Rise time is a measure of the ability of inverter virtual inertia to respond to a frequency event. The virtual inertia response rise time (10%-90%) for the control diagram in Figure 2 is determined by its parameters:

$$T_{r(10\%-90\%)} = ln9 \cdot \sqrt{T_{lpwi}^2 + T_{lwoi}^2} \approx 2.197 \cdot \sqrt{T_{lpwi}^2 + T_{lwoi}^2} \quad (6)$$

### C. Maximum steady-state response ROCOF (Hz/s).

For a certain virtual inertia time constant, the maximum steady-state ROCOF in which a PV inverter can ensure inertia power output is determined by available PV headroom. The maximum steady-state ROCOF ($ROCOF_{max}$) is determined by the headroom reserve and the steady-state virtual inertia time constant:

$$ROCOF_{max} = \frac{P_{inv\_headroom}}{2 \cdot H_{inv\_steady\_state} \cdot C_{inv}} \cdot f_N \quad (7)$$

It can be noted that for a certain level of headroom, the steady-state inverter virtual inertia time constant is inversely proportional to $ROCOF_{max}$. Substitute (5) into (7), $ROCOF_{max}$ can also be expressed as a function of the headroom reserve and controller parameters.

$$ROCOF_{max} = \frac{P_{inv\_headroom}}{K_i \cdot T_{woi1} \cdot C_{inv}} \cdot f_N \quad (8)$$

## IV. BACKGROUND: FREQUENCY RESPONSE AND HIGH PV PENETRATION ERCOT MODELS

### A. High PV ERCOT Models

The contingency magnitude for frequency response study in the ERCOT system is 2.75 GW generation trip (two units in the South Texas Nuclear Power Plant), as defined as the Resource Contingency Criteria by NERC. The ERCOT system frequency response for RCC in different renewable penetration scenarios is shown in Figure 3. It can be seen that the frequency will cross under-frequency load shedding line when the renewable penetration rate increases up to 40%. (It's noted that frequency responsive load was not simulated in this situation to study the impact of renewable penetration increase.)

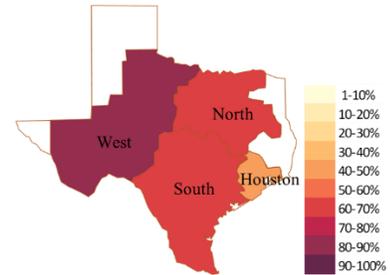

Fig. 1. ERCOT PV penetration rates in difference market zones

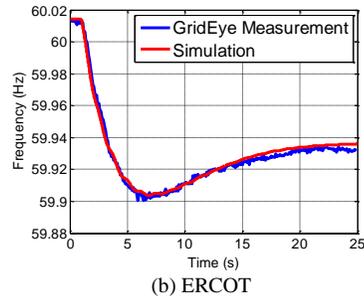

(b) ERCOT

Fig. 2. ERCOT model validation (measurement and simulation results)

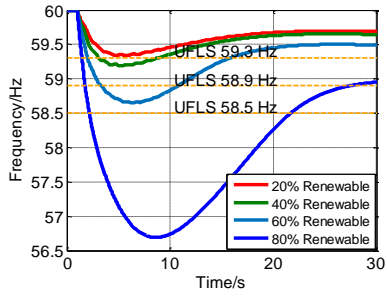

Fig. 3. ERCOT frequency response for RCC contingencies

## V. PRIMARY FREQUENCY AND AGC CONTROL OF PV

Figure 4 shows the frequency droop control diagram. It includes a deadband, a low-pass filter, a control gain, and a magnitude limiter. The low-pass filter can filter out the noises. The frequency droop control can be applied in combination with the inertia control, as shown in Figure 5.

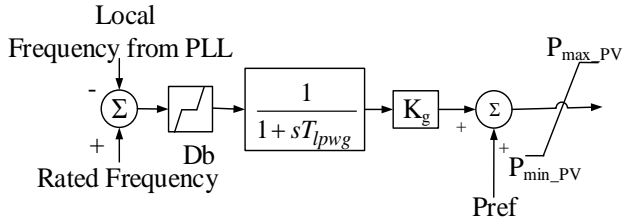

Figure 4. Frequency droop control using PV

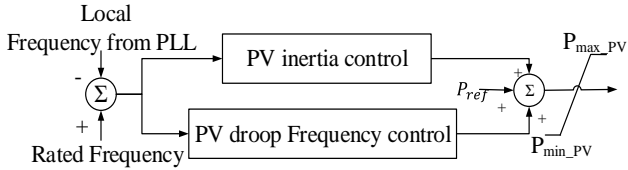

Figure 5. Primary frequency control using PV

Four control scenarios are implemented and simulated based on the high PV ERCOT model to test their effectiveness., as shown in Table 1.

Table 1. Control scenarios

| Control scenarios # | Description |
|---|---|
| 1 | PV inertia control |
| 2 | PV governor control |
| 3 | PV inertia + PV governor control |
| 4 | Without frequency control of PV |

Table2 shows the parameters in the inertia control. The control effects in the 40% renewable (65%PV + 15% WT) is shown in Figure 9. The output of PV inverters is shown in Figure 10. It can be seen that inertia control responds faster than droop control. The inertia control can delay the nadir occurrence time, whereas the droop control can increase the settling frequency. A combination of the two controls will obtain better control effects than each individual control approach.

Table 2. Parameters in the PV inertia and governor control based on frequency measurement in the EI

| Parameter | Value |
|---|---|
| Kg | 15, if enabled; 0, if not enabled. |
| Ki | 500, if enabled; 0, if not enabled. |
| Tlpwg | 1.0 s |
| Twowg1 | 1.0 s |
| Twowg2 | 1.0 s |
| Tlpwi | 1.0 s |
| Twowi | 0.1s |

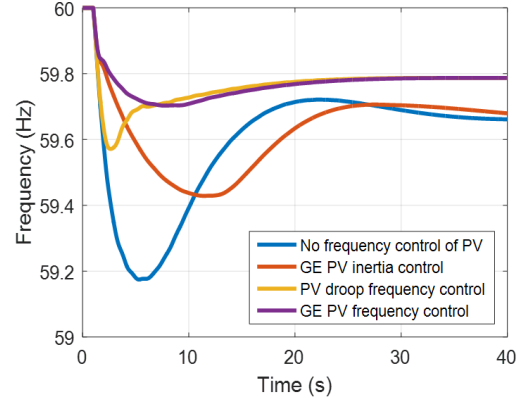

Figure 6. System frequency response using different PV control strategies in the ERCOT (40% renewable)

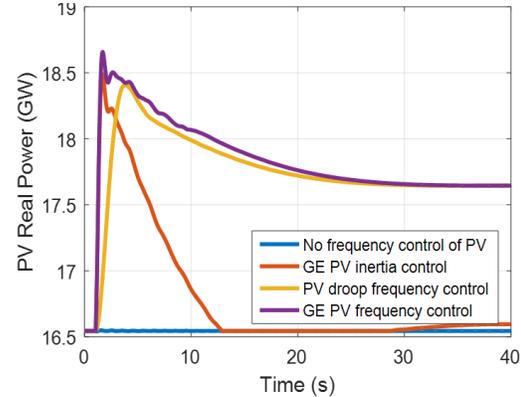

Figure 7. PV output using different frequency control strategies in the ERCOT (40% renewable)

## VI. PRIMARY FREQUENCY AND AGC CONTROL OF PV

With power reserve, PV farms can participate into AGC control. Figure 4 shows the diagram of using PV to provide AGC service. The frequency deviation signals is multiplied by a bias value and added by the inter-BA exchange power. Then the ACE signal passes through a PI link to control the real power output of PV. The control effects are shown in Figure 8 and Figure 9. Figure 8 shows the control effects with and without AGC control. Figure 9 shows the real power output of PV. It can be seen that the AGC control can bring the system frequency back to 60 Hz after it is enabled.

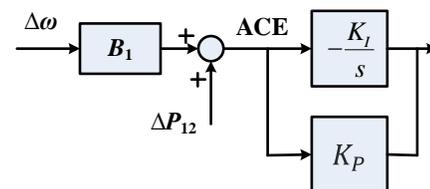

Fig. 4. AGC control diagram using PV
(Difference compared with conventional AGC control: Faster)

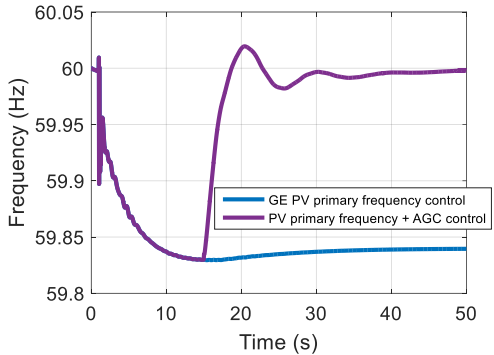

Figure 8. System frequency response with and without AGC control

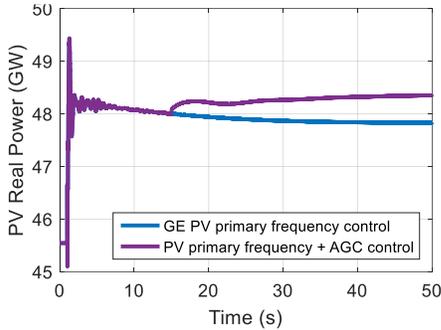

Figure 9. PV real power with and without AGC control

## VII. Fast Primary Frequency Control of PV

From Figure 8, it can be seen that AGC control has the potential to increase the frequency nadir if AGC control is enabled early after the contingency. To maximize the contribution of PV headroom, an additional control link that has similar characteristics of AGC can be added to the primary frequency control. Figure 10 shows the control diagram of this modified primary control diagram, which is called Fast Primary Frequency Control. In this control approach, an integral link is added to the primary frequency control, so that the PV headroom can be better utilized to increase the system nadir. In addition, a deadband is applied to prevent the control conflict between multiple PV farms, especially when measurement and control errors exist.

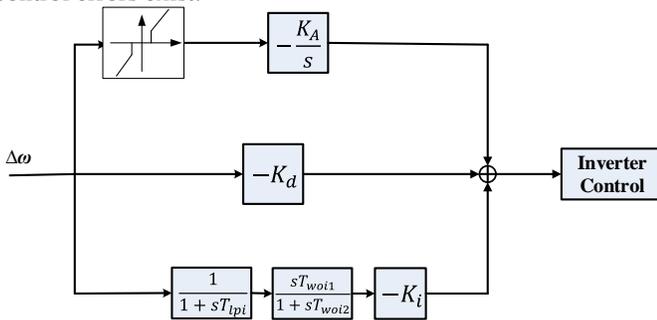

Figure 10. Fast primary frequency control using PV

Figure 11 shows the system frequency change when the Fast Primary Frequency Control is applied. It can be seen that the integral link in the control can effectively leverage the capability of PV farms to support system frequency. Figure 12 shows the real power output of PV.

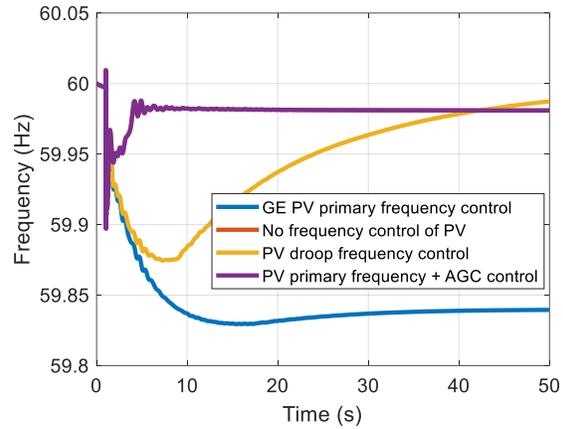

Figure 11. Frequency response with different controls

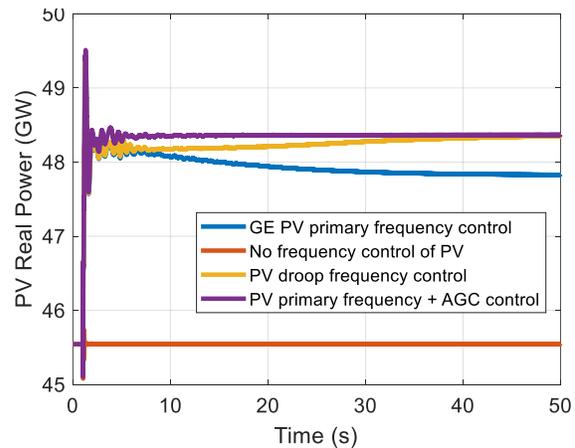

Figure 12. PV farm outputs with different controls

To exam the effects of the deadband, Figure16 and Figure 17 show the real power of PV farms whether the deadband is applied to the Fast Primary Frequency Control or not. It can be seen that the deadband can reduce the conflicts of multiple PV farms at the presence of measurement and control errors.

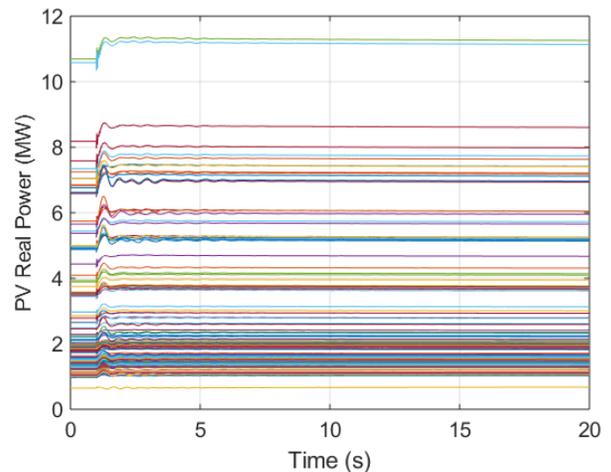

Figure 13. The real power of all PV when the integral link has a deadband

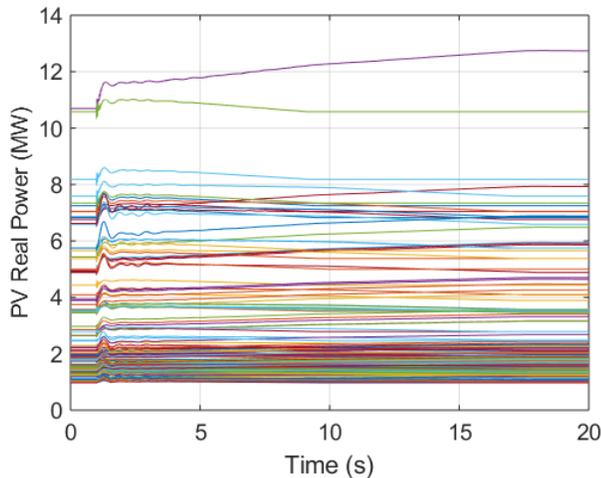
Figure 14. The real power of all PV when the integral link does not have a deadband

## VIII. Conclusions

To improve the system frequency response under high PV penetration, this paper studied different control strategies of PV farms to support grid frequency. These control include, synthetic inertia control, synthetic governor control, and AGC control. These control approaches were applied into the high PV model of ERCOT to evaluate their effectiveness. The study results show that PV has potential to provide grid frequency support that are comparable to or even stronger than conventional synchronous generators. This paper also proposed a fast primary frequency control using PV to further increase the frequency nadir after contingencies. It is noted that PV headroom reserve is required to realize these control. Future work will be optimize the PV headroom reserve based on system needs to minimize its impact on the economics of PV.
## Acknowledgements

## Acknowledgements

The authors want to express their great appreciations to all Technical Review Committee members from the following organizations: NERC, ISO-NE, ERCOT, MISO, Southern Company, TVA, PJM, Dominion Energy, NYISO, PG&E, SCE, EPRI, Peak Reliability, CAISO, and FPL, for their valuable suggestions and comments.